Groups and the Entropy Floor- *XMM-Newton* Observations of Two Groups

R. Mushotzky[1], E. Figueroa-Feliciano[1], M. Loewenstein[1,2], S. L. Snowden[1,3]
1 Laboratory for High energy Astrophysics code 662 Goddard Space Flight Center Greenbelt Maryland 20901
2 Also Department of Astronomy University of Maryland College Park MD 20742
3 Also University Space research Association

Abstract:
Using *XMM-Newton* spatially resolved X-ray imaging spectroscopy we obtain the temperature, density, entropy, gas mass, and total mass profiles for two groups of galaxies out to ~0.3 $R_{vir}$ ($R_{vir}$, the virial radius). Our density profiles agree well with those derived previously, and the temperature data are broadly consistent with previous results but are considerably more precise. Both of these groups are at the mass scale of $2 \times 10^{13}$ $M_\odot$ but have rather different properties. They have considerably lower gas mass fractions at r<0.3 $R_{vir}$ than the rich clusters. NGC2563, one of the least luminous groups for its X-ray temperature, has a very low gas mass fraction of ~0.004 inside 0.1 $R_{vir}$, which rises with radius. NGC4325, one of the most luminous groups at the same average temperature, has a higher gas mass fraction of 0.02. The entropy profiles and the absolute values of the entropy as a function of virial radius also differ, with NGC4325 having a value of ~100 keV cm$^{-2}$ and NGC2563 a value of ~300 keV cm$^{-2}$ at r~0.1 $R_{vir}$. For both groups the profiles rise monotonically with radius and there is no sign of an entropy "floor". These results are inconsistent with pre-heating scenarios which have been developed to explain the entropy floor in groups but are broadly consistent with models of structure formation which include the effects of heating and/or the cooling of the gas. The total entropy in these systems provides a strong constraint on all models of galaxy and group formation, and on the poorly defined feedback process which controls the transformation of gas into stars and thus the formation of structure in the universe.

Introduction
It has long been known (Kaiser 1986) that the form of the X-ray temperature /luminosity relation for clusters is too steep to be explained by self similar cluster evolution which only includes the effects of gravity. Kaiser (1991) and Evrard & Henry (1991) pointed out that the observed relation between cluster X-ray luminosity and temperature, $L(x) \sim T^3$ (Horner et al. 2002), could be obtained if clusters accreted pre-heated gas. This idea has received support from the recent analysis of *ROSAT* and *ASCA* observations (Ponman et al. 1999; Lloyd-Davies et al. 2000; Loewenstein 2000; Finoguenov et al. 2002) which show that the central regions of groups of galaxies have considerably more entropy than can be accounted for by shocks during the group formation (Eke et al. 1998). This "extra" entropy seemed to be roughly constant from object to object at the group mass scale and becomes relatively less important for more massive systems.

The presence of this extra heat is presumably directly connected to the need for "feedback" (Kaufmann et al. 1993; Kay et al. 2002) in semi-analytic models of galaxy formation, which is necessary to produce the observed structure of the universe. While the luminosity function of galaxies is one of the possible relics of this feedback process (van den Bosch 2002), the "extra" heat in groups and clusters should be a direct measure of this process. That is, the X-ray emission from groups and clusters *"catches"* the feedback in the act and probes its effective strength. Understanding the origin, nature, and amount of this cosmic "feedback" is a major effort in understanding cosmological structure formation and one of the main unsolved problems in large scale structure formation models. As discussed by Menci & Cavaliere (2000), it is this feedback process which dominates the amount and the thermal state of all baryons in the shallow potential wells that form at high redshifts, where most of the stars that dominate the mass of early type galaxies are created. This process thus governs both the X-ray emission from clusters and groups and the star formation rate of the universe over a wide range of redshifts. The largest uncertainties in galaxy formation models are related to the relative efficiency of cooling, feedback, and star formation, and how these regulate what fraction of the baryons are converted into luminous matter, what fraction is ejected out of the dark matter haloes by feedback processes, what fraction remains in the hot X-ray emitting phase, and where and how the metals are distributed (van den Bosch 2002).

There have been many ideas to explain the origin of this "excess" entropy including the effects of supernova heating (Brighenti & Mathews 2001; Loewenstein 2000), pre-heating of the intergalactic medium at high redshift (Tozzi & Norman 2001), the effects of quasars (Wu et al. 2000; Valageas & Silk 1999), and the importance of cooling in the gas (Bryan 2000). Recent detailed numerical studies of cluster and group formation (Borgani et al. 2001; Balogh et al. 2001; Babul et al. 2002; Brighenti & Mathews 2001) confirm that many of the correlations between cluster temperature, luminosity, entropy and mass can be explained by one of three mechanisms: 1) the accretion of high entropy gas at moderate to high redshift, 2) the heating of gas inside the bound system, or 3) the effects of cooling in a cosmological model (Muanwong et al. 2002). Several authors (Valergas & Silk 1999; Wu et al. 2000; Borgani et al. 2001) find that the observed correlations are more difficult to explain by internal heating due to supernova because of assumptions about the total energy available from reasonable supernova rates and the efficiency of the conversion of supernova shock energy into heat. However recent observations of galactic winds in rapidly star forming galaxies (Strickland 2001) indicate that most of the supernova energy goes into mass motion and heat rather than being radiated away, thus mitigating some of these criticisms. Of course the combination of both heating and cooling (Menci & Cavaliere 2000; Voit et al. 2002) is probably the most physically reasonable of all the scenarios, but is the most difficult to simulate properly.

All the numerical and analytic studies show that the accretion of high entropy gas pre-heated gas must produce an isentropic core (Balogh et al. 1999; Brighenti & Mathews 2001) which dominates the intracluster medium at the mass scale of groups ($\sim 10^{13}$ M$_\odot$). Thus a fundamental test of the entire concept of pre-heating is the determination of the entropy distribution in groups and then a direct comparison with numerical models. This has proven difficult prior to the launch of *XMM-Newton* and *Chandra* because of the low fluxes of groups, the relatively poor angular resolution of the *ASCA* telescope and the considerable controversy over the handling of the complex *ASCA* point spread function, and the low spectral resolution of the *ROSAT* PSPC. In addition there has been disagreement in the literature (cf. Buote 2000) on the actual measurements of temperature and abundance at the group mass scale because of the malleability of the X-ray CCD data and the lack of understanding of the nature of the cooling flows in the central regions and their effects on the overall *ASCA* and *ROSAT* spectra. There is extensive literature on the temperature and density of groups from the *ROSAT* satellite (e.g., Davis et al. 1996; Lloyd-Davies et al. 2002) which indicates that the temperature profiles are in general rather flat (but see Ponman & Bertram 1993 for HCG62 which shows a drop with radius), which combined with the well determined density profiles, indicates a rising entropy profile in the groups, inconsistent with preheating. However the *ROSAT* data, with their low spectral resolution, were subject to modeling uncertainties and in some cases seem to disagree with the *ASCA* temperatures (Hwang et al. 1999).

We have obtained high signal to noise X-ray spectra with *XMM-Newton* (Jansen et al. 2001) of two groups, NGC4325 and NGC2563. These objects were chosen because they span the full range in the log L (x) vs T (x) plane (Horner et al. 2002) at T~ 1 keV and have very different surface brightness profiles. Thus these two objects should probably be representative of the widest possible range of physical parameters at the group mass scale. We are able to verify that the temperatures and abundances derived with the *ASCA* X-ray CCD spectral imaging data, are in good agreement with the *XMM-Newton* EPIC data and the high spectral resolution but lower signal to noise RGS data for the central regions of NGC4325. This agreement essentially eliminates the uncertainty in spectral modeling discussed by Buote (2000). The *XMM-Newton* PSF and the very high signal to noise CCD X-ray spectroscopy have allowed a determination of the temperature, density, and chemical composition profile out to ~0.3 $R_{vir}$ and a resolution of the central cool regions allowing a precise determination of the radial entropy profile. We find that this profile is relatively steep in sharp contradiction with pre-heating models for objects at the mass scale of these groups M (500)~$4 \times 10^{13}$ $M_\odot$. These data also allow the measurement of the chemical composition of the gas in the group, the gas fraction and the total binding mass. We use a distance of 89.6 $h_{50}^{-1}$ Mpc for NGC2563 and 155.8 $h_{50}^{-1}$ Mpc for NGC4325, with effective virial radii of 1.22 and 1.44 $h_{50}^{-1}$ Mpc respectively.

Observations
*XMM-Newton* observed NGC2563 on 2001 October 15 for ~21 ks (observation ID 0108860501) and NGC4325 on 2000 December 24 for ~21 ks (observation ID 0108860101). We extracted the spectra using the standard SAS software and constructed response functions for the MOS and the PN appropriate for the extraction regions used. The original exposure depth was chosen to allow the determination of the temperature and density profiles out to ~2/3 of the virial radius, but the higher than anticipated *XMM-Newton* background has limited the range over which one can reliably determine these parameters at the present time. We hope that improvements in the modeling of the *XMM-Newton* background will allow extension of the analysis to larger length scales.

The data were screened for periods of proton flaring which required the exclusion of some exposure from the NGC4325 observation but not from the observation of NGC2563. Background spectra were collected in an annulus outside of the radius where the object radial profiles reached background levels (480-960" for NGC2563 and 420-600" for NGC4325. We chose to use local background because of a mismatch of the archival background files at low energies with the data. This is probably due to differing values of the soft X-ray background in these regions from those used for the background files. There are also problems at E>3 keV in the background subtraction, but these do not effect the presently derived temperatures, densities and entropies. The use of the "standard" background files provided by the XMM GOF requires a renormalization of the background at either low energies or high energies to match the observed spectra in the outermost rings. Since the background in XMM consists of several components it is not clear that a simple renomalization process is the right thing to do . Given the present state of the XMM background modeling and the relative insensitivity of our results to the background we fell that our present procedure is the best on can do at present. Point sources were removed from both the source and background.

SAS was used to create both redistribution matrices and ancillary region files (RMFs and ARFs) for the spectral fitting. The data we analyzed in annular rings for these highly symmetric objects. While there is some evidence for localized spectral structure in the X-ray color images (Mushotzky et al. in preparation) the effects are small.

The RGS data were analyzed with version 5.3 of the SAS. While the extended nature of the source in the RGS requires, in principle, a more detailed analysis (Peterson et al. 2002), simulations performed by the RGS team show that the standard extraction with direct allowance for the one dimensional source size (Rasmussen 2002) allows a relatively robust estimate of the temperature and abundance.

The density profiles are derived from the surface brightness profile in the 0.3-2 keV band and the use of the deprojection model in XSPEC. The derived density is quite consistent with the *ROSAT* values published by Helsdon & Ponman (2000) and Trinchieri et al. (1997). The β model parameters are listed in Table 1.

TABLE 1

| Object | r(core)arc sec | β | r(core w/o Gaussian) | β (with Gaussian) |
|---|---|---|---|---|
| NGC4325 | 21.5+/-0.5 | 0.56 +/-0.01 | 19.3+/-1.0 | 0.53+/-0.01 |
| NGC2563 | 1 | 0.34 +/-0.01 | 5.0+/-1.3 | 0.37+/-0.02 |

A better fit to both sources is obtained ($\Delta\chi^2 = 42$ for NGC4325 and $\Delta\chi^2 = 113$ for NGC2563 with the addition of a Gaussian component to the surface brightness. In NGC2563 the existence of a separate surface brightness peak in the center is quite obvious and statistically significant. These "extra" surface brightness components correspond roughly to the optical size of the central galaxy and, in the hardness ratio images, the regions where the temperature starts to drop. The X-ray image of the NGC4325 is above background out to a radius of ~7' and NGC2563 out to a radius of ~8'.

We have extracted spectra in rings for both objects (Table 2) and fitted them with the APEC model (Smith et al. 2001) with variable abundance and free column density in each ring in the 0.3-3.5 keV band . There is very little signal from the thermal gas at higher energies. The Columbia RGS team (Xu et al. 2002) has found for the spectra of Capella and the giant elliptical galaxy NGC4636 that the APEC model is an excellent overall description of the thermal spectra but that there are still remaining problems with certain Fe lines. While these objects both have effective temperatures of ~0.6 keV, and thus are not perfect analogs for the somewhat hotter NGC4325 and NGC2563, we believe that the APEC model represents the most accurate determination of the state of the hot plasma. The derived temperatures are very similar to those obtained from the older MEKAL code. (e.g., in the 30-60" ring the APEC and MEKAL temperatures differ by <1.5%). For plasmas in the 0.6-1.5 keV range with CCD X-ray data, the derived temperature is almost completely controlled by the effective energy of 2-3 Fe L line blends, and thus is relatively robust to calibration uncertainties. However, the derived temperatures are very sensitive to the Fe ionic balance. The RGS data for NGC4325 provide independent confirmation since the derived temperature depends on Fe L line ratios from Fe XVII-Fe XXIV. For the 1' extraction used in NGC4325, a compromise between a large enough area to obtain sufficient signal and a small enough area to limit the degradation of spectral resolution, the projected temperature is 0.82+/-0.02 keV compared to the EPIC average of 0.80 +/-0.02 keV, in excellent agreement. The Fe abundance in the RGS data lies between 0.2-0.8 solar at 90% confidence, in agreement with the EPIC results of 0.32 solar. We do not find evidence for multi-phase gas, nor for the near solar abundances indicated by Buote's 2001 analysis. Analysis of the RGS NGC 4325 data with a 2 component model reduced $\chi^2$ by only 2.3 which is not significant for the 2 additional degrees of freedom.

NGC4325

In the *ROSAT* PSPC data and *XMM-Newton* PN data the source is detected out to a radius of ~400-500" before the flux in the source model is consistent with the background. At a distance of 160 $H_{50}$ Mpc, this corresponds to 375 kpc or ~60% of R(500) (cf. Helson & Ponman 2001).

NGC2563

NGC 2563 has a lower overall surface brightness than NGC4325 and it is not entirely clear where the *XMM-Newton* data reach the background level. However, because of the low surface brightness, the temperature data are background limited at large radii. In the angular range from 1-12' the surface brightness distribution is well fit by a pure power law of slope –0.7+/-0.1, indicating a very flat density distribution.

Analysis:

We used the XSPEC "deproj" to obtain the deprojected deprojected radial distribution of. Using $H_o=50$, we calculated the diameter and volume of each ring, and taking the normalization of the APEC model, we calculated the density profile

,

where D is the distance to the object in cm, V is the volume of the ring, K is the APEC normalization, and µ is the ratio of helium density to electron density and is equal to 1.2. From these two quantities the entropy is readily calculated since

$$S = \frac{T}{n_e^{2/3}}.$$

Errors were propagated from the spectral fits to the entropy.

The temperature profiles were then fit empirically with the lognormal function:

$$T = T_o + A\exp\left[-\left(\frac{\ln(r/r_o)}{w}\right)^2\right].$$

The density was fit with beta models, and from these two analytical functions we calculated the total integrated mass:

$$M_{tot}(<r) = -\frac{k_B}{G\mu m_p}TR_{virial}r\left(\frac{d\log n_e}{d\log r} + \frac{d\log T}{d\log r}\right)$$

where $r=R/R_{virial}$. The gas mass was found integrating the density:

$$M_{gas} = 4\pi\mu m_p \int n_e r^2 dr,$$

and the cooling time was calculated from

$$t_{cool} = \frac{3}{2}\frac{k_B T}{n_e \Lambda}$$

where Λ is taken from Sutherland and Dopita,!1993..

NGC4325

We show in Figure 1 and Table 2 the deprojected profiles. We note that inside of 80" (10"=77 kpc) the gas temperature drops slowly, reaching a minimum value of ~0.72 keV. At larger radii the temperature is almost isothermal out to ~300" with perhaps a drop at r >300" with the outermost bin at r~0.25 $R_{vir}$ being ~85% of the peak values. The value of the temperature in the outermost bin is sensitive to the manner in which background is subtracted and the systematic exceeds the statistical error by a factor of 3.

The surface brightness is well described by the beta model with a central gas density of $1.8 \times 10^{-2}$ cm$^{-3}$. The previous analysis of the *ROSAT* temperature profile (Helsdon & Ponman 2000) had very large errors at R>1' and the nature of the temperature profile was unclear. The *XMM-Newton* temperature data are in reasonable agreement with the *ROSAT* results and the density data are in excellent agreement.

The average abundance of 0.35 solar for the single temperature model is in good agreement with the values of Davis et al. (1999), as is the average temperature when fitted to the same model. The asymptotic temperature of 1.0 keV agrees very well with the *ASCA* average temperature of 1.06 keV found by Davis et al. (1999) and 0.98 keV found by Finoguenov et al. (2001).

Our results do not support the multiphase models of Buote (2000) for this object, and the disagreement is discussed in detail in our subsequent paper. However, the range of temperatures seen in the *XMM-Newton* data nicely bracket the two-temperature solution found by Buote (2000). In the central regions the cooling time is less than $10^9$ y at radii less than ~50 kpc, almost exactly where the gas temperature starts to drop. One of disturbing characteristics of the data, and what may have driven the multiphase conclusion of Buote (2000), is the presence of absorption in excess of the galactic column ($2.2 \times 10^{20}$ cm$^{-2}$). This is seen in both the EPIC (MOS and PN) and RGS data, and thus we believe it is real. This value which peaks at ~$6 \times 10^{20}$ cm$^{-2}$ in the group center, slowly drops with radius reaching a galactic value at R>200". However because the exact value of this absorption depends sensitively on background subtraction at low energies one has to take this result with some caution. The effect on the temperature is very small.

The imaging analysis shows that the temperature and abundance, with one small perturbation, is remarkably symmetric in this object and supports the use of simple circular extraction radii

Table 2 for NGC 2563

| Radius [arcs] | Temperature [keV] | $n_e$ [cm$^{-3}$] | S [keV cm$^2$] | Abundance |
|---|---|---|---|---|
| 30 | 0.77 | 7.97E-03 | 19.30 | 0.26+/-0.06 |
| 60 | 1.27 | 1.77E-03 | 86.70 | 0.29+/-0.14 |
| 120 | 1.55 | 1.01E-03 | 154.03 | 0.29+/-0.13 |
| 240 | 1.6 | 4.90E-04 | 257.41 | 0.20+/-0.05 |
| 480 | 1.35 | 3.83E-04 | 256.05 | 0.18+/0.03 |
| 960 | 1.02 | 1.53E-04 | 355.82 | 0.15+/-0.02 |

Table 2 for NGC 4325

| Radius [arcs] | Temperature [keV] | $n_e$ [cm$^{-3}$] | S [keV cm$^2$] | Abundance |
|---|---|---|---|---|
| 15 | 0.72 | 2.09E-02 | 9.50 | 0.36+/-0.05 |
| 30 | 0.749 | 1.53E-02 | 12.13 | 0.29+/-0.03 |
| 60 | 0.819 | 6.17E-03 | 24.35 | 0.34+/-0.04 |
| 120 | 0.982 | 2.00E-03 | 61.78 | 0.28+/-0.05 |
| 180 | 1.033 | 1.12E-03 | 95.58 | 0.20+/-0.04 |
| 240 | 1.01 | 5.29E-04 | 154.37 | 0.14+/-0.03 |
| 360 | 0.955 | 5.24E-04 | 146.90 | Undetermined |

## The abundance is with respect to Anders and Grevesse 1989

NGC2563

Like NGC4325, this object is remarkably circular. The color image shows that the gas temperature starts to drop at at r<10 kpc almost exactly the optical stellar radius and where the cooling time is less than $10^9$ y. However, as opposed to NGC4325, this radius is well within the central galaxy. There is no evidence, like in all the other *XMM-Newton* cooling flow spectra, of multi-phase gas. The low central surface brightness inside 30" did not allow a RGS spectrum with sufficient signal to noise to derive sensible limits on the temperature

The *XMM-Newton* data for this object are in reasonable agreement with the temperature profiles derived from the *ROSAT* data (Trinchieri et al. 1997) but with a systematic offset of ~10%. This is consistent with what is seen in other *ROSAT* and CCD measurements (Hwang et al. 1999). The overall fit to the deprojected data has $\chi_\nu^2 = 1.20$ fitting over the 0.3-3.5 keV band, with the bulk of the excess $\chi^2$ contribution at E>2 keV in the outer rings, indicating a small problem with background subtraction. However, the model never deviates by more than 10% from the data.

III Entropy
   A. Results

Consistent with the previous *ASCA* (Finoguenov et al. 2002) and *ROSAT* analysis (Trinchieri et al. 1997; Helsdon & Ponman 2000), we find an almost isothermal temperature profile outside of the central regions, which combined with a dropping density indicates a rising entropy profile. The vastly improved signal to noise relative to *ROSAT* and spatial resolution relative to *ASCA* have allowed us for the first time to precisely measure the entropy profiles. The entropy profiles scaled to 0.1 of the virial radius are different with S=120 keV cm$^{-2}$ for NGC4325 and S= 280 keV cm$^{-2}$ for NGC2563, with essentially no statistical error. Thus our data do not support the concept of an entropy floor which is constant from object to object, nor the existence of a uniform pre-heated intergalactic medium. In addition the very similar abundances indicate that the metals left behind in the gas phase do not correlate with the entropy. Furthermore the low gas fraction of NGC2563 combined with its similar average abundances compared to NGC4325 indicates that the energy associated with the observed metals is less in NGC2563 compared to NGC4325 yet it has twice as much entropy.

However, despite having very similar average temperatures (1.36 keV and 0.95 keV for NGC2563 and NGC4325, respectively), abundances, and masses, the values of the entropy at similar $R/R_{vir}$ are different and the shape of the entropy profiles are also different. These values of the entropy at R~0.1 $R_{vir}$ are very similar to other groups as found by Lloyd-Davies et al. (2000) and Finoguenov et al. (2002). The much higher *XMM-Newton* signal to noise and better angular resolution allows us to see for the first time that both the absolute values and the shapes of the entropy profiles differ from object to object at the same mass scale.

As shown by Muanwong et al. (2002), the major differences in the entropy distribution between the pre-heating and cooling models is in the inner regions, in particular at $R<0.2R_{vir}$, which have been well measured by the present results. At large radii the pre-heating models can either resemble adiabatic models (see also Balogh,!Babul,!and Patton 1999) for similar results) or, depending on the degree of pre-heating, can even be isentropic. At large radii cooling models give similar results to pure adiabatic models.

IV Masses and gas fractions

The total masses derived from the *XMM-Newton* spectroscopic imaging data are in very good agreement with the observed scaling laws from higher temperature systems (Horner 2002) and support the use of the average X-ray temperature as a reasonable surrogate for mass. Despite their rather different X-ray surface brightness distributions, and somewhat different average temperatures outside of the central .01 $R_{vir}$, the total mass profiles of the two systems is extremely similar. However the gas mass fractions profiles are radically different (figure 2). NGC2563 is very gas poor in the central regions, similar to that seen in many elliptical galaxies (e.g., NGC4636, Loewenstein & Mushotzky 2002 ) while NGC4325 has a much flatter gas mass fraction profile. However, at ~$0.3R_{vir}$ the gas mass fractions are similar at ~4-6% of the total mass, which still lies much below the average value for rich clusters (~16% with our adopted Hubble constant of 50 km/sec/Mpc, Allen, Schmidt, & Fabian 2002). We believe that this argues strongly for some process that either removes the gas from the central regions or pushes it further out. However, because the residual gas has similar metal abundances this process must not change the average gas abundance of the remaining gas.

V Discussion:

As shown in detail in Babul et al. (2002) and Voit et al. (2002), preheating models require that the entropy profile be very flat out to large radii where shocks begin to dominate. In NGC2563 the entropy profile steepens at R~30" or ~0.02 of the virial radius, much smaller than allowed at this mass scale for preheating models , for NGC4325 the entropy flattening occurs at a similar radius. We thus conclude that the present data are inconsistent with pre-heating models. In addition pre-heating models would predict that at this mass scale the total entropy would only have a small contribution from shocks and thus that the entropy of systems with similar mass should be identical, contrary to our observations.

The question then arises, are these groups unique or are they representative of the general status of groups? In the survey of Horner et al. (2002), NGC4325 is one of the most luminous groups for its X-ray temperature and thus, given its "normal" structural parameters (its β value of 0.58 is almost exactly the average that Helsdon & Ponman 2000 and Mulchaey et al. 2002a find for groups with kT~0.7-1.0 keV), it must have a higher than average gas density at this mass range. In the mass-temperature relation of Finguenov et al. (2001), NGC4325 lies slightly above the best fit regression line at the group mass scale but exactly on the best fit extension from massive clusters. Like many other groups NGC4325 has a very small number of bright galaxies; however, the velocity dispersion can be very well measured from the large number of small galaxies in these systems (Zabludoff & Mulchaey 1998). NGC4325's ratio of temperature to velocity dispersion ($\sigma$ =265+/-50 km s$^{-1}$) is one sigma below the best fit regression between temperature and velocity dispersion. Thus despite its high X-ray luminosity NGC4325 shows no obvious anomalies in its properties.

NGC2563 on the other hand is one of the least luminous groups for its temperature and its surface brightness profile is extremely flat. Both of these facts are perfectly consistent with its higher entropy at r~0.1 $R_{vir}$ and its shallower entropy profile. However its total mass is not unusual for its temperature and the relationship between X-ray temperature and optical velocity dispersion ($\sigma$ = 336 + 44, -40 km s$^{-1}$) is also not unusual.

We thus conclude that, contrary to much of the previous work, (but in agreement with Brighenti & Mathews 1999 analysis of NGC 4472) that there is not a uniform entropy floor in groups but that there is a range of entropies.

The origin of this variance in entropy is not certain but it might reflect a range of histories in the cooling or heating for the groups. As shown in Borgani et al. (2001) and Brighenti & Mathews (2001), internal heating models are consistent with these entropy profiles. In the internal heating scenario, the energy input corresponding to the "extra" entropy depends sensitively on the relative history of the object, e.g., when and where the heat was produced relative to the collapse epoch of the object. Thus in these models one might expect to have a range of central entropies and entropy profiles.. The observed range of X-ray luminosities at a fixed temperature (Babul et al. 2001) is then due to the wide range of collapse times of objects in the mass range from $5 \times 10^{12}$ M$_\odot$ to $5 \times 10^{12}$ M$_\odot$ .

In the cooling scenarios the apparent range of entropy and entropy profiles (Davé, Katz, & Weinberg 2002) presumably is due to the range of densities present over the wide range of collapse times and thus the variable effects of cooling. We note that the calculations of Davé et al. (2002) seem to reproduce the properties of NGC2563 extremely well, in density, entropy, and luminosity at the observed temperature scale. This is consistent with the fact that nowhere in NGC2563 is the density high enough (outside the central galaxy) for the gas to be cooling with a cooling time less than $3 \times 10^9$ yr. However the Davé et al. (2002) calculations do not produce an object like NGC4325, with its high density and high luminosity and lower entropy. Inspection of the predicted $L(x)$ vs T diagrams in the Davé et al. (2002) paper does not show objects with the high X-ray luminosity of NGC4325. The predicted range of entropies from the cooling models of Muanwong et al. (2002) seems to match the observed scatter. There have not yet been similar calculations for the internal heating models.

Given the apparent ability of the internal heating, cooling, and cooling+heating (Voit et al. 2002) scenarios to grossly explain the data, one must look for more detailed tests of the models. All three of these scenarios have a lower gas density for groups than "adiabatic" gravitational collapse models in agreement with observations. A possibly observable distinction between the "cooling only" and "heating only" scenarios is the relative amount of cooled gas, that presumably has formed stars. One thus might expect a relative decline in the hot gas mass fraction of groups at the same relative virial radius compared to clusters and a relative increase in the mass of stars. The radical difference in the gas mass fraction for NGC2563 in the central regions seems to be inconsistent with its very similar total stellar mass to NGC4325 (Mulchaey et al. 2002b) if cooling dominates the entropy origin. We thus tentatively favor models in which internal heating dominates the creation of the "extra" entropy but that the effects of cooling are also important.

Other Tests of the Models:
Davé et al. (2002) point out that their models show similar density laws over a wide range of masses, apparently in disagreement with the data. The observed *XMM-Newton* surface brightness profiles out to R~0.3 $R_{vir}$ are well fit by a two component model with β (fit) and R(core) being well determined for NGC4325 but with R(core) poorly determined for NGC2563. This indicates that there is a real difference in surface brightness laws, in apparent disagreement with the Davé et al. (2002) calculations. However, with only two objects so far one can say little about trends, except that the very flat surface brightness seen in NGC2563 has never been seen in a massive, rich cluster.

We believe that our higher accuracy *XMM-Newton* profiles support the general reliability of the *ROSAT* analysis and thus confrim the wide variety of surface brightness profiles, at least out to ~0.3 $R_{vir}$, which again gives some support to the internal heating models.

Scatter in the L (x) vs T plane:
As shown by the two representative objects in this paper and the much larger sample of *ASCA* clusters and groups (Horner et al. 2002), there is a wide range of scatter in the L(x) vs T plane which increases at lower mass scales. From inspection of the numerical models such scatter also exists, but is not discussed in detail in the papers. It is now clear from these two objects that the scatter is not due to gross differences in metallicity or to morphology or mass in stars (e.g., the integrated galaxy luminosities), but to entropy differences and true differences in the gas density and temperature profiles.

Both of these objects have very low values of β (spec)= 0.52 for NGC2563 and 0.42 for NGC4325 based on precise X-ray temperatures and well sampled optical velocity dispersions (Zabludoff & Mulchaey 1998). These low values agree with the simulations of Davé et al. (2002) at these mass scales. Muanwong et al. (2002) compare the specific energy of the gas to that of the dark matter at the relevant mass scale. They find that the inclusion of cooling raises the ratio of energy in the gas to the dark matter by up to 40%, consistent with our values of β (spec). However, Davé et al. (2002) seem to find steeper values of β (fit) than our observed ones, however Muanwong et al. (2002) find values quite consistent with our observations.

The global temperature profiles are rather different in the two groups. NGC2563's profile seems to be in rather good agreement with all the recent theoretical predictions, (e.g., the universal profile of Loken et al. 2002) but NGC4325 is rather flatter in its profile. However, the temperature drop in the center seen in both of these objects does not appear in many of these models. In the heating calculation of Brighenti & Mathews (2001), there is not a direct comparison of the gas vs dark matter temperature nor of the X-ray surface brightness models, and thus we are unable to comment on this type of comparison.

Both the pre-heating and cooling calculations of Muanwong et al. (2002) find a wide range in predicted entropy at ~0.1 $R_{vir}$ at the same mass scale, consistent with the present observations. The fraction of the baryonic mass that is in gas in the theoretical calculations of Muanwong et al. (2002) are presented at the virial radius and thus are very difficult to compare with our data, since the total mass estimates are very uncertain out to such large radii. The baryonic fraction in hot gas is also calculated by Davé et al. (2002) and they find a very strong variation in fraction of the baryons in the hot gas with X-ray temperature, and at $<T>\sim 1$ keV this value is $\sim 0.2-0.4$, down by $\sim 50\%$ from that in more massive clusters. This is roughly consistent with the values we find for NGC4325 but very different from the low baryonic fraction in NGC2563 at $\sim 0.1 R_{vir}$. However, the ratio of the baryons that lie in the gas varies strongly with radius in NGC2563, and thus comparison with the models is difficult.

The observed chemical abundances in these systems are consistent with the values reported in Davis et al. (1998), and do not show a strong radial gradient. Thus the variance in entropy is not directly correlated with the chemical abundance. If the heat which produces the "extra" entropy has been provided by supernova, then most of the heavy elements have either been ejected from the group or pushed to larger radii, and are not observed in the present data. The preliminary indication of a drop in abundance in both of these systems at large radii (Table 2) favors ejection and thus the enrichment of the IGM, in a scenario similar to that proposed by Davis et al. (1998).

Recently Borgani et al. (2002) have run a simulation in which both the effects of heating due to supernovae and cooling are taken into account at the mass scale of our groups. The entropy profile and values that they find are in good agreement with our data for NGC4325 but lie well below NGC2563. As they point out, they have underestimated the cooling due to the neglect of line radiation, which dominates at these temperatures and which would tend to increase the entropy. However their simulation suffers from the over-cooling problems frequently seen in such calculations and thus it is not clear how reliable these results are.

A clue to the relevant important process is the change in the apparent conversion of gas into stars, which occurs at the group mass scale (Marinoni & Hudson 2002). As they point out the poor-group mass is the scale at which the mass-to-light ratio of virialized systems begins to increase which suggests a physical link between star formation and the X-ray properties of halos. Similar ideas have been suggested by Bryan (2000) who formulated them as the efficiency of galaxy formation being higher in groups than in clusters. However this has been strongly criticized by Balogh et al. (2001).

Conclusions:

We have shown for the first time that the entropy of groups at the same mass scale and at the same virial radius can differ and that there is does not exist a universal entropy floor. The position of the two groups analyzed here in the L(x) vs T plot at the extrema indicates that we may have sampled the full entropy range at this mass scale of roughly a factor of two. The entropy profile of these two systems also differs subtly.

The entropy profiles and the lack of a constant entropy value rule out pre-heating models (Borgani et al. 2002; Babul et al. (2002); Tozzi & Norman 2001). The entropy profiles are broadly consistent with those calculated in both radiative cooling (Davé et al. 2002; Muanwong et al. 2002) and internal heating (Brighenti & Mathews 2001) models. Internal heating models must have a consistent source of heat since the observed entropy range is rather small and thus if the heat source is due to AGN it must be very finely tuned to the cluster potential, while if it is due to supernovae the total amount of energy seems larger than normally assumed, but maybe consistent with what is required in semi-analytic models (Kay et al. 2001). It is interesting that in the Kay et al. (2001) model the history of the particles that they have chosen as representative ends up with an entropy very similar to that of the particles at r ~0.1 $R_{vir}$ in our groups.

The pure cooling models seem to require too much of the hot gas to cool out into cold baryons compared to the observed stellar mass in the galaxies (Mulchaey et al. 2002a). It thus seems likely that both heating and cooling are important (Voit et al. 2002). The absence of a precise entropy floor indicates that the processes which sets the minimum entropy has cosmic scatter, but only on the order of 50%. Based on the simulations we speculate that this scatter is due to the combination of a wide range of formation redshifts for objects at the group mass scale (Babul et al. 2000) combined with the epoch of star formation. In this scenario one might get different entropies for systems in which the stars in the galaxy form before the group collapses, compared to systems in which the group collapses first. Of course there could be considerable additional heating due to active galaxies which could easily add some variance. In a companion paper (Mulchaey et al. 2003), we compare the mass fractions of cooled gas in stars and hot gas compared to the virial masses, which shows that the fraction of gas that has cooled is not a strong function of mass scale.

Given the large variance in apparent entropy profiles we would like to stress the risk of extrapolation of the models beyond the range of observation. The low baryonic fraction of NGC2563 completely disappears at r>0.3 $R_{vir}$, and given the shallow surface brightness may completely reverse sign. Such cautions also apply to the abundance data. It is now apparent that the data for the low surface brightness, low luminosity, flat β(fit) groups are highly biased relative to the more "normal" groups and that comparison of their global properties requires great care.

These data provide strong tests of the "feedback" parameter on galaxy formation models; unfortunately these models have not been presently formulated to predict how much heat is injected into the gas (Kay et al. 2002). However, these simulations need the maximum amount of heating allowed by supernova (Kay et al. 2002) to provide a good match to the galaxy luminosity function, which is very similar to the values inferred from the X-ray entropy data and suggests that all the "extra" entropy is due to supernova heating. Direct comparison of the ensemble of entropy profiles for many groups combined with detailed optical data will strongly constrain all galaxy formation models and might even allow age dating of these systems.

Future *XMM-Newton* observations will allow detailed entropy profiles for many more groups out R~0.3 $R_{vir}$ but given the higher than anticipated background levels it will be very difficult to obtain high quality temperature profiles at larger radii. It will require a much lower background experiment with reasonable angular resolution to go to larger radii; this should be possible with *Astro-E2*.

Figure Captions
Figure 1
Panel a: The deprojected temperature profile for NGC2563 and NGC4325 vs. virial radius. The horizontal error bars represents the width of the annular bins used in the analysis.
Panel b: The deprojected density profile for NGC2563 and NGC4325 vs. virial radius.

Panel c : The entropy profile for NGC2563 and NGC4325 vs. virial radius

Figure 2

Panel a: Encircled mass vs radius for NGC2563 and NGC4325 vs. virial radius. The horizontal error bars represent the width of the annular bins used in the analysis.

Panel b: Encircled gas mass vs radius for NGC2563 and NGC4325 vs. virial radius

Panel c : The ratio of gas mass to total mass for NGC2563 and NGC4325 vs. virial radius

Acknowledgements: We thank M. Still for extraction of the RGS data. We thank the XMM project for the provision of such excellent data.

References

Allen, S., Schmidt, R., & Fabian, A. 2002, astro-ph/0205007

Babul, A., Balogh, M. L., Lewis, G. F., & Poole, G. B. 2002, MNRAS, 330, 329

Balogh, M. L., Babul, A., & Patton, D. R. 1999, MNRAS, 307, 463

Balogh, M. L., Pearce, F. R., Bower, R. G., & Kay, S. T. 2001, MNRAS, 326, 1228

Borgani, S., Governato, F., Wadsley, J., Menci, N., Tozzi, P., Lake, G., Quinn, T., & Stadel, J. 2001, ApJL, 559, L71

Borgani S., Governato F., Wadsley J. N. P., Quinn, T., Stadel, J., & Lake, G. 2002 MNRAS, in press

Brighenti, F. & Mathews, W. G. 1999, ApJ, 512, 65

Brighenti, F. & Mathews,!W.!G. !2001, ApJ, 553, 103

Bryan, G. L. 2000, ApJ, 544, L1

Buote, D. A. 2000, MNRAS, 311, 176

Davé, R., Katz, N., & Weinberg, D. H. 2002, ApJ, submitted.

Davis, D. S., Mulchaey, J. S., Mushotzky, R. F., & Burstein, D. 1996, ApJ, 460, 601

Davis, D. S., Mulchaey, J. S., & Mushotzky, R. F. 1999, ApJ, 511, 34

Eke, V., Navaro, J., & Frenk, C. S. 1998, ApJ, 503, 569

Evrard, G. & Henry, J. P. 1991, ApJ, 383, 95

Finoguenov, A., Reiprich, T. H., & Böhringer, H. 2001, A&A, 368, 749

Finoguenov, A., Jones, C., Boehringer, H., & Ponman, T. J. 2002, ApJ, 578, in press

Helsdon, S. F., & Ponman, Trevor, J. 2000, MNRAS, 315,

Horner, D. 2001, PhD Thesis, University of Maryland

Horner, D. J., Baumgartner, W. H., Gendreau, K. C., & Mushotzky, R. F. 2002, ApJ, submitted

Hwang, U., Mushotzky, R. F., Burns, J. O., Fukazawa, Y., & White, R. A. 1999, ApJ, 516, 604

Kaiser, N. 1986, MNRAS, 222, 323

Kaiser, N. 1991, ApJ, 383,104

Kauffmann, G., White, S. D. M., & Guiderdoni, B. 1993, MNRAS, 264, 201

Kay, S., Pearce, F. R., Frenk C., & Jenkins, A. 2002, MNRAS, 330, 113

Loewenstein, M. 2000, ApJ, 532, 17

Loewenstein, M. 2001, ApJ, 557, 553

Loewenstein M., & Mushotzky, R. 2002, ApJ, submitted

Lloyd-Davies, E. J., Ponman, T. J., & Cannon, D. B. 2000, MNRAS, 315, 689

Loken, C., Norman, M., Nelson, E., Burns, J., Bryan, G., & Motl, P. 2002, astrop-ph 0207095

Marinoni, C. & Hudson, M. J. 2002, ApJ, 569, 101

Menci, N. & Cavaliere, A. 2000, MNRAS , 311, 50,

Muanwong, O., Thomas, P. A., Kay, S. T., & Pearce, F. R. 2002 MNRAS, in press

Mulchaey, J. S., Davis, D. S., Mushotzky, R. F., & Burstein, D. 2002a, ApJ, submitted

Mulchaey, J. S., et al. 2002b, in preparation

Ponman, T. J. & Bertram, D. 1993, Nature, 363, 51

Ponman, T. J., Cannon, D. B., & Navarro, J. F. 1999, Nature, 397, 135

Rasmussen, A. 2002, private communication

Smith, R. et al. 2002, APEC model

Strickland, D. 2001, astro-ph0107116

Sutherland, Ralph S. and Dopita, M. A. 1993 ApJ. Suppl vol. 88,. 253

Tozzi, P.; Norman, C. 2001, ApJ, 546, 63

Trinchieri, G., Fabbiano, G., & Kim, D.-W. 1997, A&A, 318, 361

Valageas, P. & Silk, J. 1999, A&A, 350,725

van den Bosch, F. C. 2002, MNRAS, 332, 456

Voit, G., Mark Bryan, G. L., Balogh M. L, Bower, R. G. 2002, ApJ, in press

Wu, K. K. S., Fabian, A. C., Nulsen, P. E. J. 2000, MNRAS, 318, 889

Xu, H., Kahn, S. M., Peterson, J. R., Behar, E., Paerels, F. B. S., Mushotzky, R. F., Jernigan, J. G., & Makishima, K. 2002, ApJ, submitted

Zabludoff, A. I. & Mulchaey, J. S. 1998, ApJ, 496, 39

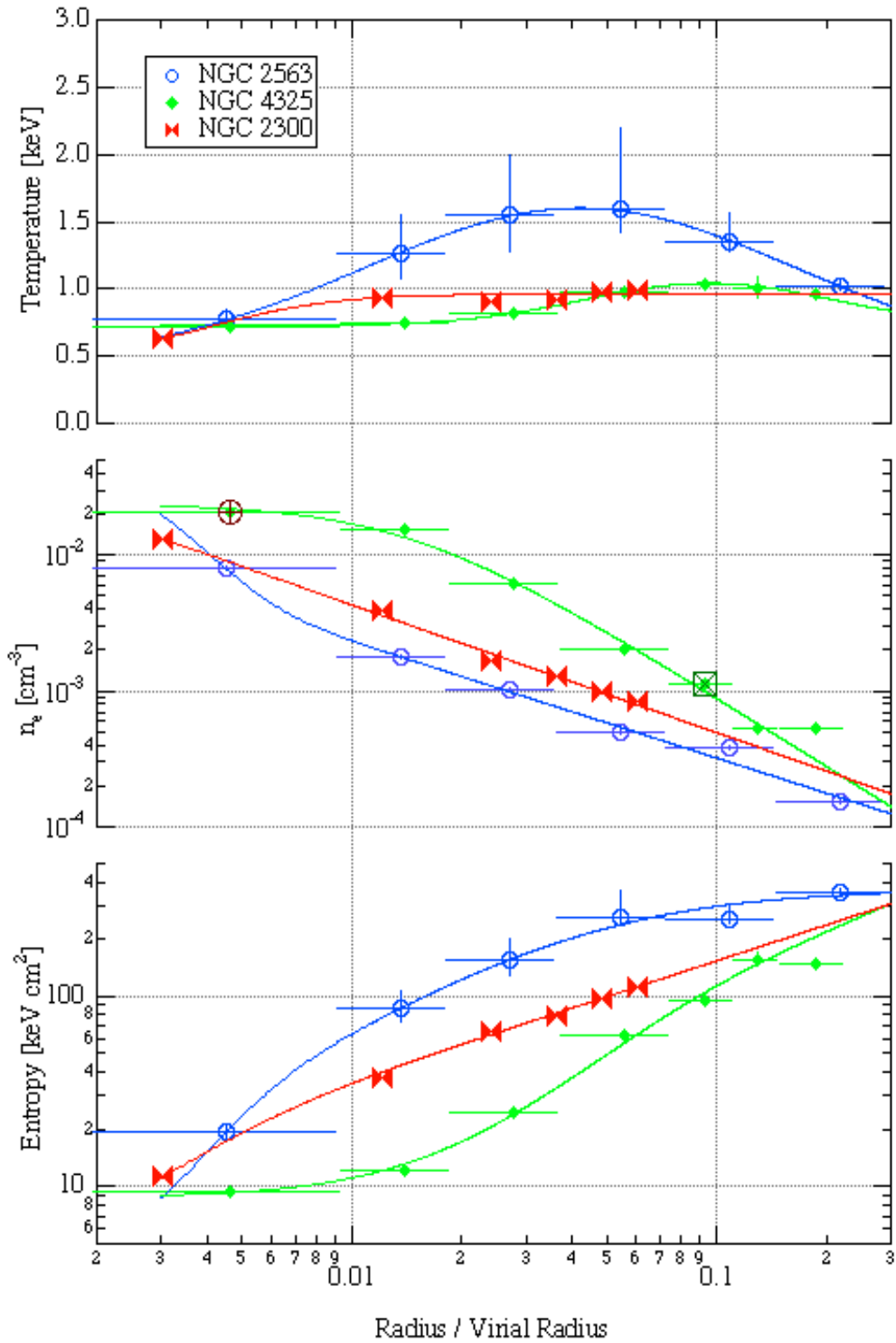

Figure 1.

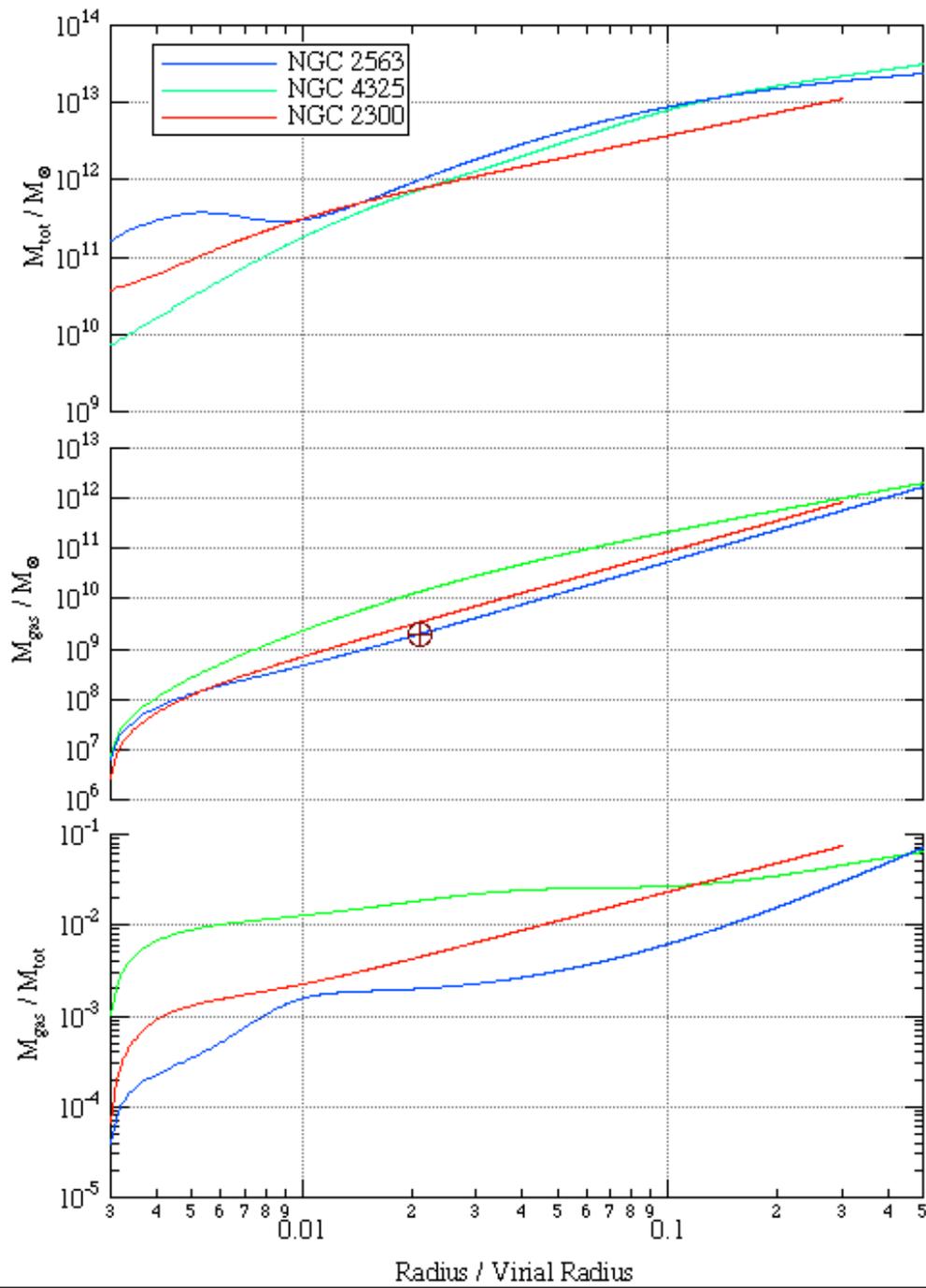

Figure 2.